\documentclass[conference]{IEEEtran}
\IEEEoverridecommandlockouts
\usepackage{cite}
\usepackage{amsmath,amssymb,amsfonts}
\usepackage{algorithmic}
\usepackage{graphicx}
\usepackage{textcomp}
\usepackage{xcolor}
\usepackage{hyperref}
\def\BibTeX{{\rm B\kern-.05em{\sc i\kern-.025em b}\kern-.08em
    T\kern-.1667em\lower.7ex\hbox{E}\kern-.125emX}}
\begin{document}

\title{Heterogeneous ALU Architecture - Power Aware System}

\makeatletter
\newcommand{\linebreakand}{%
  \end{@IEEEauthorhalign}
  \hfill\mbox{}\par
  \mbox{}\hfill\begin{@IEEEauthorhalign}
}
\makeatother

\author{\IEEEauthorblockN{Alok Anand}
\IEEEauthorblockA{\textit{Electrical \& Computer Engineering} \\
\textit{Carnegie Mellon University}\\
Pittsburgh, PA, U.S.A. 15213 \\
\href{mailto:aloka@andrew.cmu.edu}{aloka@andrew.cmu.edu}}\\
\and
\IEEEauthorblockN{Ivan Khokhlov}
\IEEEauthorblockA{\textit{Electrical \& Computer Engineering} \\
\textit{Carnegie Mellon University}\\
Pittsburgh, PA, U.S.A. 15213 \\
\href{mailto:ikhokhlo@andrew.cmu.edu}{ikhokhlo@andrew.cmu.edu}}\\
\and
\IEEEauthorblockN{Abhishek Anand}
\IEEEauthorblockA{\textit{Electrical \& Computer Engineering} \\
\textit{Carnegie Mellon University}\\
Pittsburgh, PA, U.S.A. 15213 \\
\href{mailto:aanand3@andrew.cmu.edu}{aanand3@andrew.cmu.edu}}\\
\linebreakand

}

\maketitle

\begin{abstract}
The advent of heterogeneous multi-core architectures brought with it huge benefits to energy efficiency by running programs on properly-sized cores. Modern heterogeneous multi-core systems as suggested by Artjom et al.\cite{Grudnitsky2015ARP} schedule tasks to different cores based on governors that may optimize a task for energy use or performance. This provides benefits to the system as a whole in reducing energy costs where possible, but also not compromising on performance for timing-critical applications. In the era of dark silicon, energy optimization is increasingly important, and many architectures have arisen that seek to optimize processors to specific tasks, often at the cost of generality. We propose that we can still achieve energy-saving and potentially performance-improving benefits while not affecting a system's generality at all, by achieving heterogeneity at the level of Arithmetic logic unit (ALUs). Much like a heterogeneous multi-core system achieves benefits from its heterogeneity and efficient scheduling, a heterogeneous ALU system can achieve similar benefits by routing ALU operations to properly sized ALUs. Additionally much like there are scheduling modes for the governors of heterogeneous multi-core processors, we propose that energy-constrained modes can be effective in a heterogeneous ALU system with the routing of operations to smaller ALUs for immense energy savings. We examine the energy and performance characteristics of scaling ripple carry adders and evaluate the total energy and performance benefits of such a system when running applications. With our proposed controls, input operand size-based and energy constraint-based, we could potentially emulate the success of heterogeneous processor task scheduling at a finer-grained level. This paper presents our evaluation of the potential of heterogeneous ALU processors.  
\end{abstract}

\vspace{1em}

\section{Introduction}
Presently, heterogeneous processor systems as reported in Todman et al.\cite{article} work, schedule tasks to re-configurable functional units to low power/frequency (or high power/frequency/resource) processor cores depending on performance requirements to great benefit in energy costs of operation. This work aims to replicate that success at the granularity of ALUs.

Modern processors incorporate fixed single operand size Arithmetic Logic Units (ALUs) which process all operations with relatively identical delay and energy consumption. Thus, though instruction operands for a specific ALU may vary in size throughout a program, they incur similar energy consumption and Cycles Per Instruction (CPI) cost for each invocation of that ALU.

This paper examines the characteristics of different-sized ripple-carry (RC) adders as analogous to the potential benefits of ALUs as a whole. Several configurations were evaluated to estimate performance and energy benefits based on different modes of scheduling. The two scheduling modes we considered were: inputs sent to adders that match the largest operand's size, and all inputs to one size adder, which we termed heterogeneous and homogeneous configurations respectively. The benefit of the heterogeneous configuration is estimated as performance and energy improvements, and the main benefit of homogeneous configurations is estimated as energy savings at some performance cost. 

This work suggests the potential for the creation of custom ALU-level schedulers that could power on or off different ALUs, and route operations to different ALUs as desired by the system, thus emulating CPU governors.

Synthesis of adders allowed us to gain measurements for the characteristics of these different-sized adders. Evaluation of a CPU benchmark informed about potential total benefits and costs of these system configurations. In order to do so, though we did not perform a full system integration, we counted the occurrence of different-sized operands in calls to the ADD instruction for the Dhrystone benchmark, allowing us to estimate processor operation characteristics based on the characteristics of the different adders synthesized.

\section{Motivation}
Past research on the characteristics of heterogeneous single-core computing by Chung et al. work{\cite{10.1109/MICRO.2010.36} and multi-core processors shows comparatively good performance and energy efficiency results compared to their counterparts in homogeneous multi-core processors {\cite{2}}  {\cite{1}}. If such benefits could be seen at a core-level heterogeneous implementation, this work hypothesizes that they could be seen when heterogeneity is taken to a finer granularity, looking at the ALUs inside a processor core. Targeting ALUs specifically to achieve improvements in performance and energy efficiency, when compared to processing units comprising homogeneous ALUs, could allow for the implementation of an operand-routing system much like the scheduler governors of heterogeneous processors. Cores as a whole throughout a multi-core system could be set to minimize energy costs of operation, allowed to optimize for performance, or some mix of the two based on the system's governor.

\section{Main Idea}
With this work, we address the pressing question: Do “Heterogeneous ALU processors” have potential?
Current processor core architecture, either 32-bit or 64-bit, has homogeneous-sized ALU. A program may have instructions with varying operand sizes. The homogeneous design creates a fixed energy wall for instruction based on ALU creating the same wall for a program. Running a program on a homogeneous core has a minimum energy requirement that is high compared to the energy requirement when operand size is considered during ALU operation.
The minimum energy requirement of a homogeneous core makes it unsuitable for energy-constrained systems like portable systems, and energy harvesting \cite{10.1145/3296957.3173210} and limits the working or active process time in intermittent computing devices.

Heterogeneous cores utilize the benefits of heterogeneous operand functional units, both at full and reduced available energy. Mapping instructions with ALU size in the heterogeneous core provide reduced energy consumption compared to the homogeneous ALU core. Even the source has full energy available as shown in \textit{Fig.\ref{fig:Power-aware full power}}, heterogeneous cores save more energy than homogeneous. At reduced energy, directing instruction to smaller operand ALUs in hetero-operand ALU core takes less energy and provides increased core working time and can have a major impact on decreasing inactive during intermittent operation as mentioned in Lucia et al. work\cite{lucia_et_al:LIPIcs:2017:7131}.

\begin{figure}[h!]
    \begin{center}
    \includegraphics[width=0.5\textwidth]{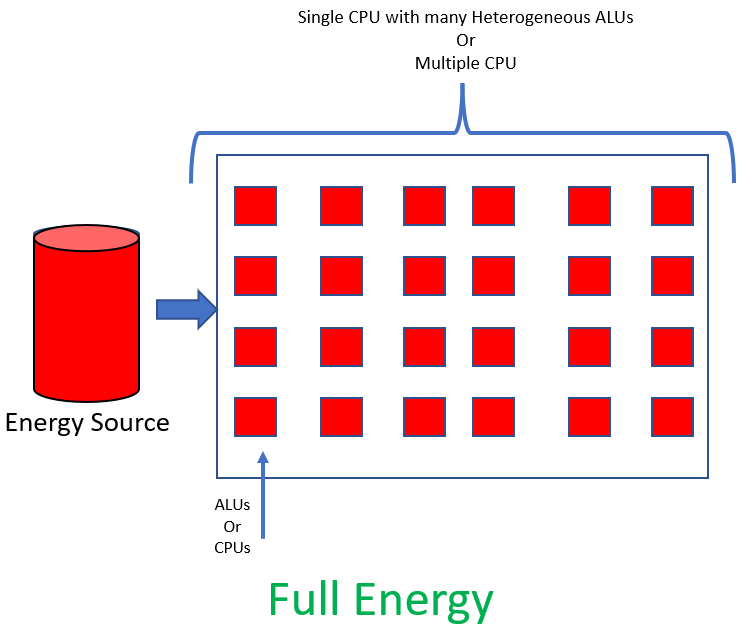}
    \end{center}
    \caption{Power aware heterogeneous architecture design  at full power}
    \label{fig:Power-aware full power}
\end{figure}

In heterogeneous processing elements design, is to have different-sized functional units such as adders executing corresponding ADD instructions owing to its operand size, heterogeneous ALU design incurs fewer energy costs, latency, and area costs, taking advantage of this design architecture would allow running CPU cores at the reduced energy level, even with full power available.
At a reduced energy level shown in \textit{Fig.\ref{fig:Power-aware reduced power}}, the heterogeneous design permits assigning tasks only to 8-bit adders, incurring overall fewer energy costs, bypassing other adders in heterogeneous design, when supplied energy level reduces from 100\% to 50\% and further to 25\% level.

\begin{figure}[h!]
    \begin{center}
    \includegraphics[width=0.5\textwidth]{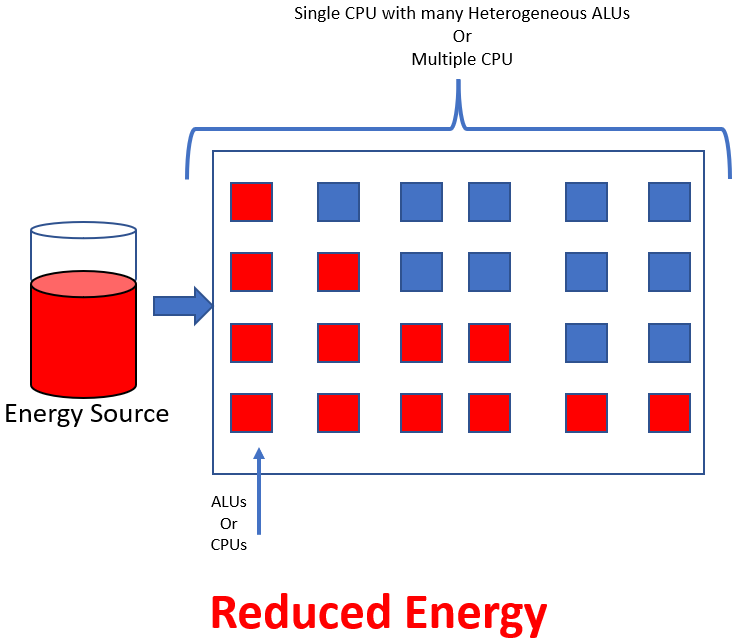}
    \end{center}
    \caption{Power aware heterogeneous architecture design  at reduced power}
    \label{fig:Power-aware reduced power}
\end{figure}

\section{Architectural Implementation}
In realization of heterogeneous operand defined ALU architecture on a CPU core as in \textit{Fig.\ref{fig: Heterogeneous ALU-ADDER design, determined according to operand size}}, this work focuses on functional unit-adder that involved writing RTL of different variants of ripple-carry adders with different operand sizes, simulating to get an estimate of energy consumption and timing traces of activity.
\begin{figure}[h!]
    \begin{center}
    \includegraphics[width=0.5\textwidth]{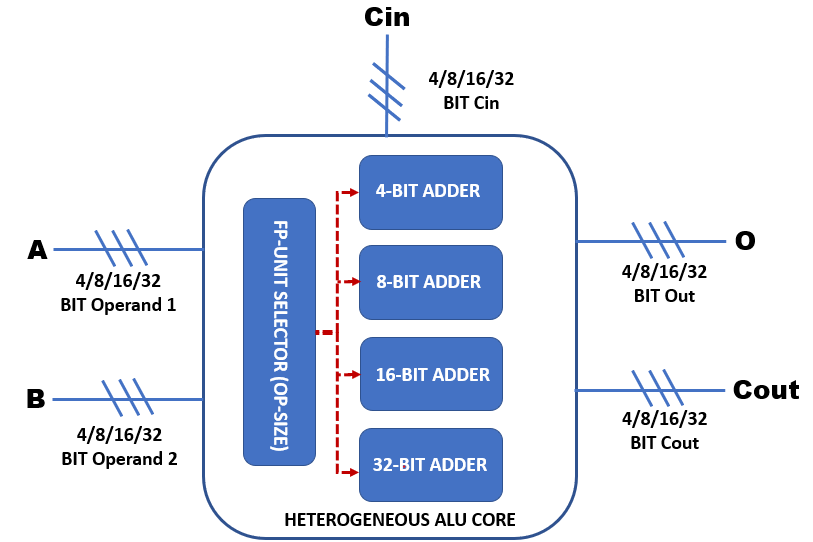}
    \end{center}
    \caption{Heterogeneous ALU-ADDER design, determined according to operand size}
    \label{fig: Heterogeneous ALU-ADDER design, determined according to operand size}
\end{figure}

In using the design of a multi-ALU system, the Design Compiler tool provides power, timing (data-arrival time), and area analyses of adder ALU designs for 4-bit, 8-bit, 16-bit, and 32-bit adders as shown with tool flow \textit{Fig.\ref{fig:adder1_8_16_32_64_energyVSoperand}}. These energy, cycle, and area costs with performed analyses of different operations performed on these adders assuming all 4 are available in the same system simultaneously.
The architecture allows analyzing 4-bit, 8-bit, 16-bit, and 32-bit addition operations on each adder, in order to determine the performance implications of different operations on different adders and attempts to determine trade-offs and optimal input-forwarding for different conditions,e.g. power and energy costs are lower for smaller adders. In case of system’s power or available energy is reduced, makes sense to perform longer cycle operations on smaller adders (though for energy specifically, optimizing for trade-offs with system static power in the future). In calculations, assumptions to have an operating frequency of 1GHz with data in conjunction with operation counts from the pin tool to determine trade-offs in latency and performance of a heterogeneous ADD ALU system in the scope of various benchmarks.
\begin{figure}[h!]
    \begin{center}
    \includegraphics[width=0.5\textwidth]{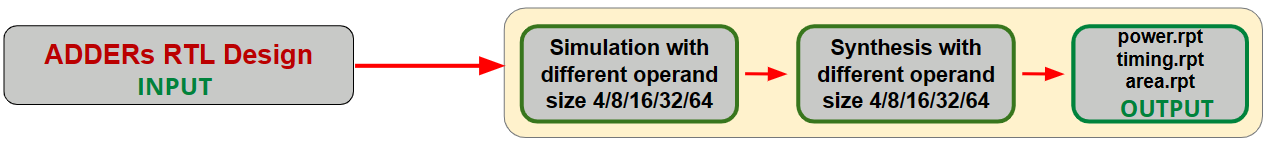}
    \end{center}
    \caption{Tool flow of design compiler, Adder simulation and synthesis to obtain power, timing, and area}
    \label{fig:adder1_8_16_32_64_energyVSoperand}
\end{figure}

\section{Methodology}
In the design of different sized verilog with ripple carry design using default '+' instructions. It includes RTL design of adders of different bit-size with the simulation of different operand sizes, including 4/8/16/32/64 bit data with each computing to produce every clock cycle. A 4-bit adder incurs 8 clock-cycle to compute and produce results with 32-Bit operand data. 
Includes the synthesis of hardware with operand size to obtain power, performance, and area report.
\begin{figure}[h!]
    \begin{center}
    \includegraphics[width=0.5\textwidth]{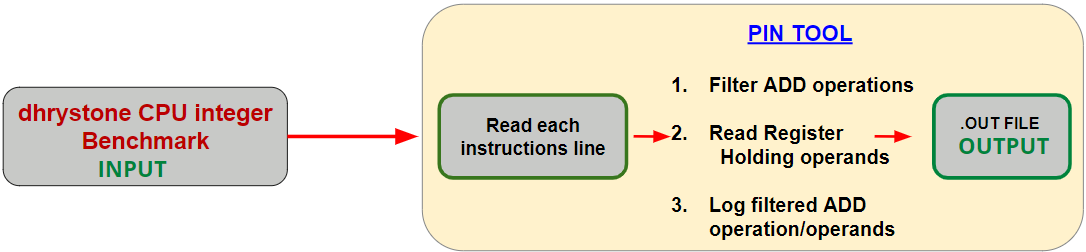}
    \end{center}
    \caption{Pin-Tool setup, running with Dhrystone CPU benchmark}
    \label{fig:Pin-Tool setup, running with dhrystone CPU benchmark}
\end{figure}

To evaluate the heterogeneous Adder ALU shown in \textit{Fig.\ref{fig:Pin-Tool setup, running with dhrystone CPU benchmark}}, we used a synthetic computing benchmark, Dhrystone "C" (integer). Since our design uses add operand, we are interested in all various operand size addition operations from the benchmark.
To extract the addition operation We used the pin\cite{3} tool.
We generated ".out" output files from Dhrystone ".C" file by compiling using GNU Compiler Collection (GCC). The generated ".out" file also includes approximately 20\% the default library instruction along with 80\% Dhrystone instruction. 
We wrote a pin tool script to filter out the instructions based on ADD opcode along with register operand value to reveal operand size. 
In order to get power and timing data for benchmarking, we need the number of ADD operations, Segregation based on operand size, and using operand count results along with energy costs and timing from experimental results to get an estimate of energy costs.
To obtain the segregated count of addition operation based on operand size, we used the output file generated after running the pin-tool script. The pin-tool output file contains instructions and registers values. In order to get the instruction count based on specific operand size within a program, we wrote a Python script that takes in the pin-tool output file and provide the instruction count. This Python script segregate instruction based on the operand value with maximum size in an instruction.
Using instruction count information, we calculate the energy cost savings and potential timing trade-offs when each instruction is executed using optimal ADDER, which means operand size and adder size is the same, in our heterogeneous operand ALU configuration. We would multiply the results of our analysis of energy costs of operations by operation counts from the pin tool.
We intend to estimate and characterize optimal ADDER combinations to perform tasks with operand size and operations counts in a program, for use in optimizing operations in an energy-constrained system.

\begin{figure}[h!]
    \begin{center}
    \includegraphics[width=0.5\textwidth]{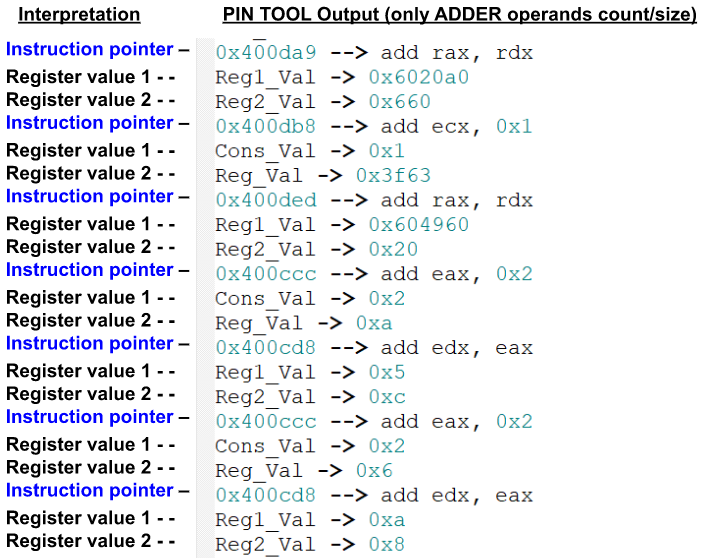}
    \end{center}
    \caption{Pin-Tool output interpretation filtering ADD operations with operands size, based on larger operand size}
    \label{fig:Pin-Tool output interpretation filtering ADD operations with operands size, based on larger operand size}
\end{figure}

\begin{figure}[h!]
    \begin{center}    \includegraphics[width=0.5\textwidth]{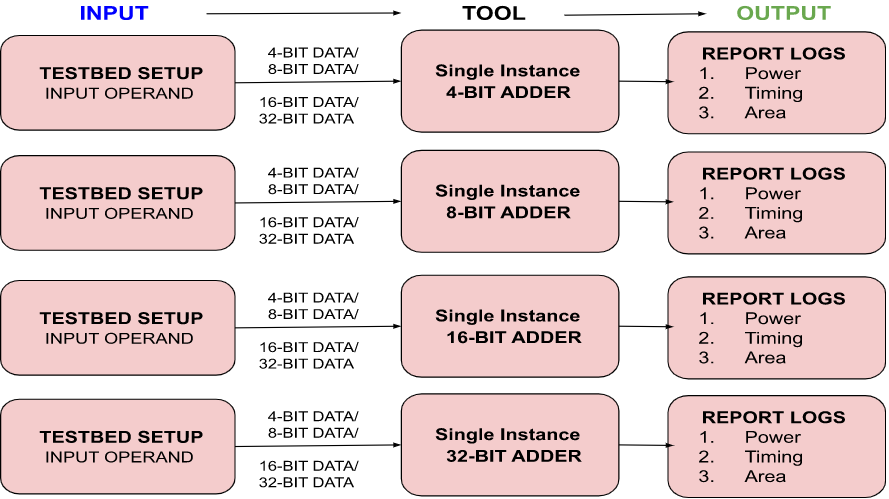}
    \end{center}
    \caption{Energy variation with N-Bit adder, executing different operand size}    \label{fig:experimentSetup}
\end{figure}

A heterogeneous adder is to perform different operand size ADD operations on different ADDER combinations as shown in \textit{Fig.\ref{fig:experimentSetup}}, instead of a homogeneous adder to get energy, timing, and area value that incurs performing one complete operation when combined with pin tool operators counts would optimal combination of Adders combination.

A homogeneous adder is to perform different operand size ADD operations on a homogeneous adder as shown in a single instance in \textit{Fig.\ref{fig:experimentSetup}}, to get energy, timing, and area value that incurs performing all complete operations on the same kind of adder combinations, and obtained result is combined pin tool operators count to compare with heterogeneous ADDER design.

\section{Experimental Design}
Experimental evaluation of the proposed design is carried out in two steps.
\subsection{Operand size characterization of ALU unit}
With pin-tool is used for placing instrumentation with each instruction within the benchmark. Inserting instrumentation after each instruction takes into account ADD operations on two memory registers, one memory with the other one as immediate data and both using immediate operands. With filtered instructions, the reading register involved in operations provides an operand that is logged in the output file. We devised a Python script to group the number of operands based on size, determined by the larger of two operands. With this data operand size and percentage, distribution showed 4-bit operations account for 19\% and operations less than 16bit account for 33\% of total decoded operand size in the program.
\subsection{Power, performance and area evaluation of ALU unit}
With the design compiler, we evaluated running different-sized operations on different-sized adders to record energy incurred, for e.g. 32-bit add operations on a 4-bit adder with increased, 8-cycle latency but with minimum energy costs.
In the Design compiler reported energy, area, and cycle time costs are combined with extracted quantities of ADD instructions of various sizes keeping all operands active same time, providing a comparison of homogeneous ALU with heterogeneous ALU design. Using design compiler data, we obtained performance characteristics and energy running entire instructions on a homogeneous system. The next steps involved calculating energy costs in case different-sized operations running on matched adder sizes reduce energy and latency overheads.  
The calculation for energy costs is expressed as: 

\textit{Energy costs = Dynamic Power consumed of adder in operation * Dynamic time of operation + Static Power of all adders * Total operation time}

\textit{where, Dynamic operation time = data arrival time * cycle time.}

Overall this is a pessimistic estimate of energy costs, as we would not expect dynamic power costs to always be maximized over all inputs, and also throughout the signal propagation period.

\begin{figure}[h!]
    \begin{center}\includegraphics[width=0.5\textwidth]{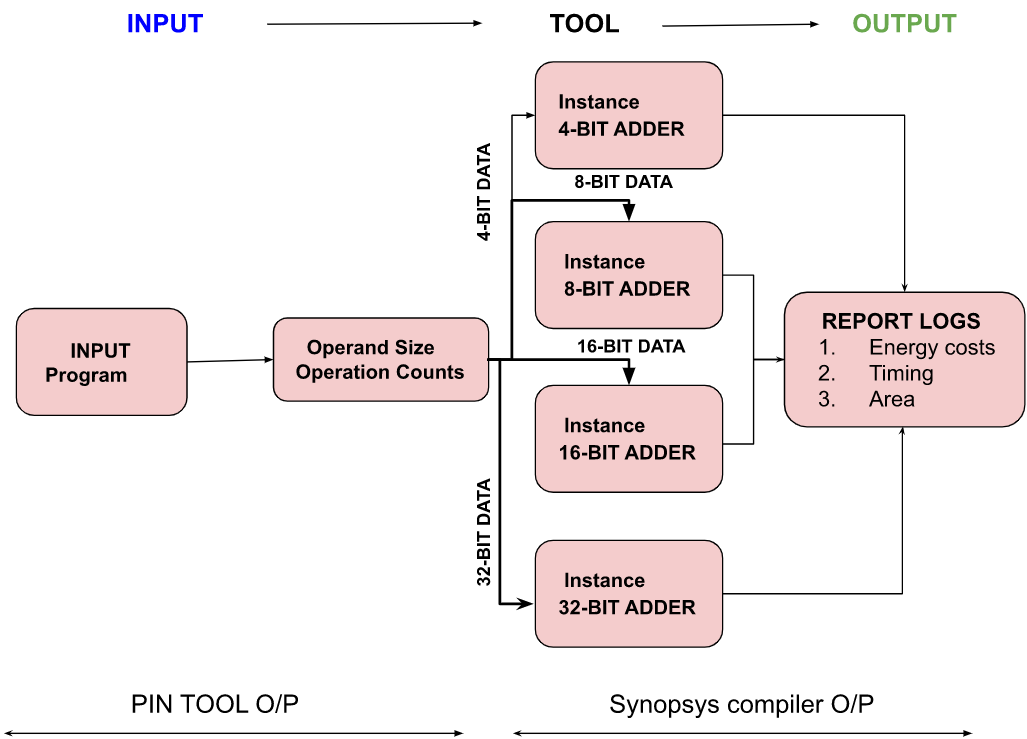}
    \end{center}
    \caption{Evaluation setup of Heterogeneous}    \label{fig:adder1_8_16_32_64_energyVSoperand}
\end{figure}

\section{Experimental Results}
\subsection{ADD Operations in Dhrystone}
After determining the occurrence of each sized ADD operation in Dhrystone, bounded on the larger of the two operands, as shown in \textit{Table.\ref{tab1}}, we disovered that 33\% of ADD operations performed in Dhrystone is of 16 bits or less. This finding is what effectively enables the heterogeneous adder configuration to have performance and energy benefits over static max-size adders, as a significant percentage of operations are able to execute more efficiently on adequately sized smaller adders. Though further analysis would be necessary, we suspect that other benchmarks and programs may have similar results for other arithmetic operations in the system. If that is the case, it may lend utility to having heterogeneous systems for other ALUs.

\begin{table}[htbp]
\caption{Dhrystone Benchmark Count of ADD Operations With N-bit Operands}
\begin{center}
\begin{tabular}{|c|c|}

\hline
\textbf{Operand Bit Size} & \textbf{Count} \\

\hline
\text{4 Bit} & \text{200,776} \\
\hline
\text{8 Bit} & \text{52,734} \\
\hline
\text{12 Bit} & \text{14,070} \\
\hline
\text{16 Bit} & \text{83,628} \\
\hline
\text{32 Bit} & \text{196} \\
\hline
\text{64 Bit} & \text{712,433} \\
\hline
\textbf{Total} & \textbf{1,063,837} \\
\hline

\end{tabular}
\label{tab1}
\end{center}
\end{table}

\subsection{Adder Characteristics}

The 8-bit adder performs well for all operations with large energy savings and minimal if any trade-offs to performance.
Looking at \textit{Fig.\ref{fig: Energy costs of an operation on data inputs of size N- bits performed on each of the adders.}} for energy costs, and \textit{Fig.\ref{fig: CPI of an ADD operation on data inputs of size N performed on each of the adders. }} for CPI costs: Both 4-bit and 8-bit operands, the optimal adders are respectively the 4-bit and 8-bit adders. However for 16-bit operands, the optimal adder was found to be the 8-bit adder, which performs the addition in the same number of clock cycles as the 16-bit adder, but for less energy. And for 32-bit operands, we have found that by using an 8-bit adder, we increase the clock cycles required to process the data to 4 as compared to 3 when using a 32-bit adder, a 33\% increase, but reduce energy costs considerably from 0.218pJ to 0.0618pJ: a 71\% decrease. Additionally, both the 16-bit and 8-bit adder take the same amount of clock cycles to process 32-bit operands, leading to the 8-bit adder strictly outperforming the 16-bit adder on energy costs while maintaining the same processing speed. 

The energy cost difference for 4-bit operation between 4 and 8-bit adders is smaller than from 32 to 8bit; .0171pJ for the 8-bit adder to .00773pJ for the 4-bit adder: a 45\% decrease for the 4-bit over the 8 bit. From the perspective of energy costs, the 8-bit adder is most optimal across all operations and still manages to bring a small 4\% improvement over the 64-bit architecture in performance. A trade-off would have to be considered for the 32-bit architecture however, as the 8-bit adder architecture incurs a 15\% penalty to performance over the 32-bit adder architecture.

In order to process larger operands, smaller adders would need to operate for several cycles. Due to varying latencies and the selected frequency of 1GHz, we found that in some cases, smaller adders could perform at the same time as a larger adder, but with sizeable corresponding trade-offs in energy favoring the smaller adder.  In \textit{Fig.\ref{fig: CPI of an ADD operation on data inputs of size N performed on each of the adders. }}, for example, the 16-bit operation can be carried out fastest by either the 16-bit or 8-bit adders, however looking at \textit{Fig.\ref{fig: Energy costs of an operation on data inputs of size N- bits performed on each of the adders.}},  the 8-bit adder would use less energy. Thus the 8-bit adder becomes an optimal choice for these operations. This occurs as few adders incur latencies closer to a multiple of a whole clock cycle, and others may be further away requiring the wait of an additional clock pulse to synchronize data (e.g. an adder that has a latency equivalent to 1.9 clock cycles needs 2 clock cycles to finish an operation, as does one that has a latency equivalent to 1.1 clock cycles since it misses the previous positive edge and its data will not be read until later). This effect is of note as it causes some adders to have strictly better performance and energy characteristics, but also as these adders maintain such characteristics when needing to operate over multiple iterations on operand inputs larger than the adder itself. That is to say, if two adders have a relationship of equal performance but one uses less energy, that relationship is maintained regardless of what size inputs are used, and in fact, the smaller adder may get a performance boost when it comes to operands of its size.

We suspect that with increased frequencies, this effect would become less pronounced, with fewer adders sharing identical CPI due to a higher granularity of clock pulses; however, there is potential to take advantage of this effect of frequency masking away the performance costs of smaller ALUs to create more energy-optimal systems with minimal to no costs to performance. We believe there would be other ALU designs alongwith frequency combinations that can effectively hide added delay within the granularity of the clock frequency, and thus provide for significant energy optimizations to simpler, smaller ALUs.

In this system an 8-bit adder would use $1/8^{th}$ of the area of the 64-bit adder, we expect this ratio to be much lower when using more complicated adder designs due to increased complexity and branching of larger designs.

\begin{figure}[h!]
    \begin{center}    \includegraphics[width=0.5\textwidth]{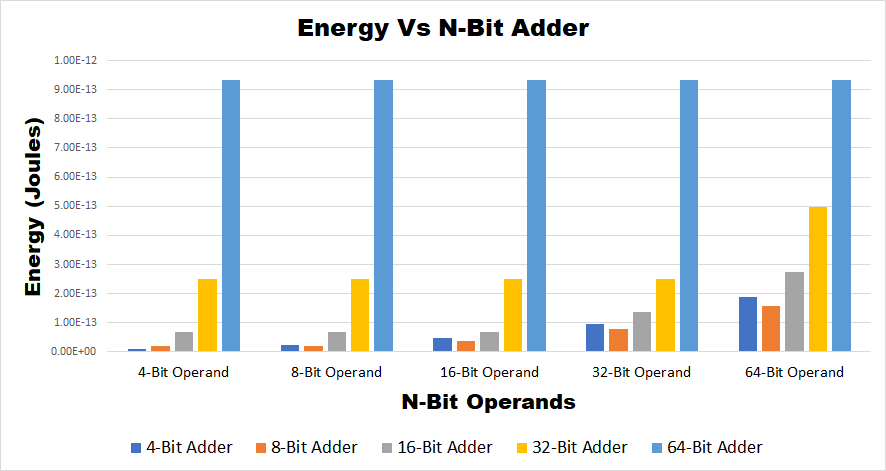}
    \end{center}
    \caption{Energy costs of an ADD operation on data inputs of size N performed on each of the adders. Smaller adders would need to operate for several cycles to add larger-bit operands.}    \label{fig: Energy costs of an operation on data inputs of size N- bits performed on each of the adders.}
\end{figure}

\begin{figure}[h!]
    \begin{center}    \includegraphics[width=0.5\textwidth]{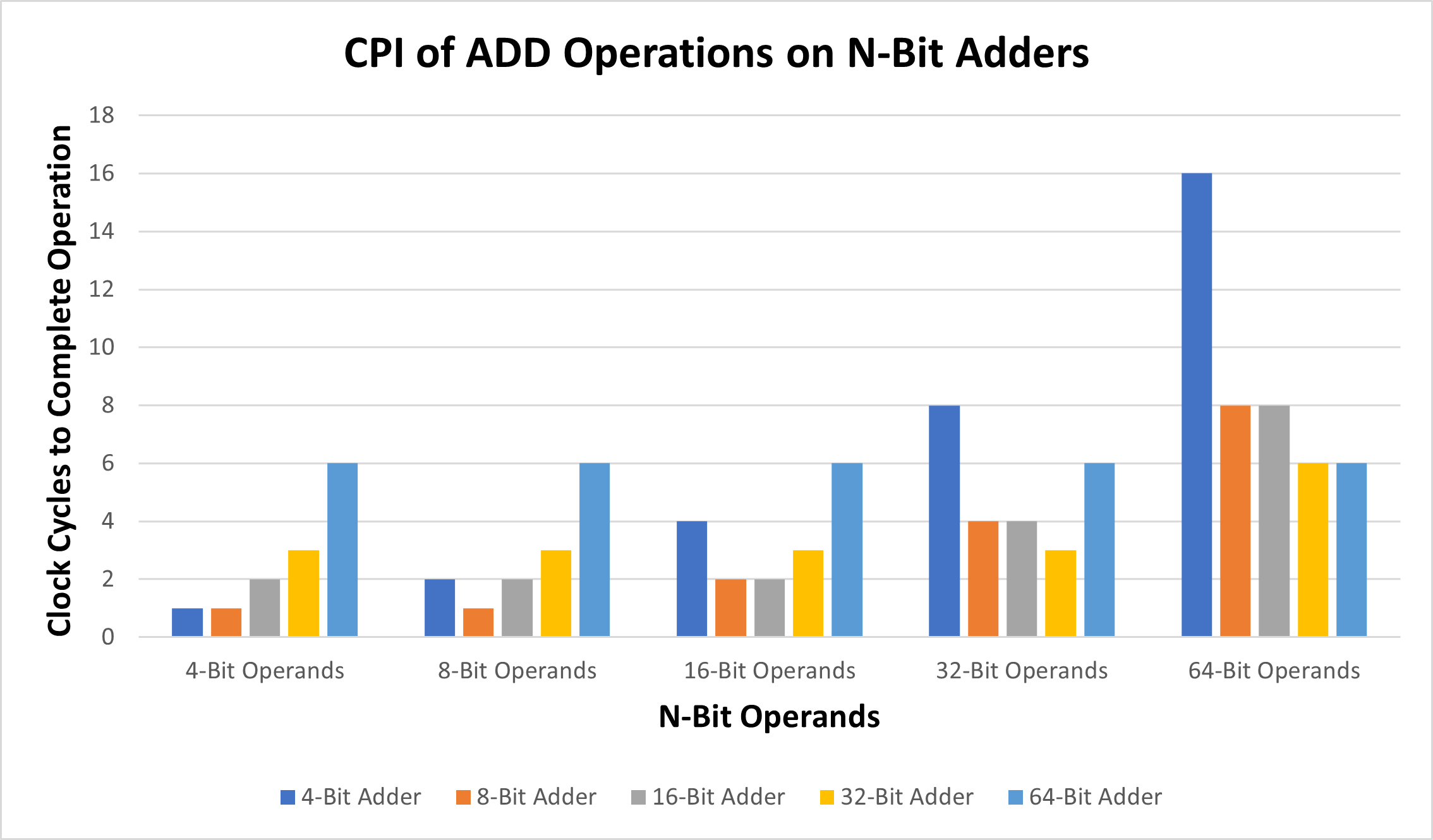}
    \end{center}
    \caption{CPI of an ADD operation on data inputs of size N performed on each of the adders. }    \label{fig: CPI of an ADD operation on data inputs of size N performed on each of the adders. }
\end{figure}

\subsection{Homogeneous ALU configuration}
Echoing our results from the characteristics analysis of the adders, the 8-bit adder performs well in both 32 and 64-bit architectures, with large energy savings and minimal if any trade-offs to performance. Additionally, the 32-bit adder is the best-in-performance for both architectures, with the cost of increased energy per instruction.

In both the 64-bit and 32-bit architectures, the 8-bit adder showed the greatest energy savings out of all adder configurations including heterogeneous. When compared to the 64-bit homogeneous system, the 8-bit adder showed a nearly 4\% performance improvement \textit{Fig.\ref{fig:normalizedCycleArch64}} and more than a 90\% reduction in energy costs \textit{Fig.\ref{fig:energy_costs_homo_hertero_64alu_arch}}. The performance of the 8-bit adder homogeneous system was about nearly 15\% worse than that of the 32-bit adder system \textit{Fig.\ref{fig:normalizedCycleArch32}}, but with nearly 80\% reduction in energy costs \textit{Fig.\ref{fig:energy_costs_homo_hertero_32alu_arch}}. 

In both the 64-bit and 32-bit architectures, the 32-bit adder showed the greatest homogeneous performance. When compared to the 64-bit homogeneous system, the 32-bit adder saw a nearly 17\% performance improvement \textit{Fig.\ref{fig:normalizedCycleArch64}} and more than a 75\% reduction in energy costs. \textit{Fig.\ref{fig:energy_costs_homo_hertero_64alu_arch}}. In the 32-bit homogeneous system, the 32-bit adder had the best performance of all adders \textit{Fig.\ref{fig:normalizedCycleArch32}}, but with greater than 80\% increased energy costs over the next largest costs with the 16-bit adder \textit{Fig.\ref{fig:energy_costs_homo_hertero_32alu_arch}}. 

Either 8-bit or 32-bit adder configurations could be useful, with the main trade-off for the two being that of energy favoring the 8-bit adder system, and performance favoring the 32-bit adder system.

In regards to potential area costs, since RC adders scale linearly, we can conclude that a homogeneous system using only 1 size of an adder will generally maintain or reduce total area utilization as any other larger adder. Compared to utilizing an adder of size X, one could use double the adders of size X/2 and maintain the same area costs. However, using double the adders may not be necessary, and maintaining the same quantity of adders results in an area reduction of 50\% instead. Optimal quantities of adders would need to be analyzed in a system-level context to account for stalls and frequencies of operations, however since adders like the 8-bit and 32-bit showed performance improvements in some circumstances due to the aforementioned frequency masking, it would likely not be necessary to increase the quantities of adders when switching to those configurations, resulting in overall area savings

\subsection{Heterogeneous ALU configuration}
In the 64-bit architecture as shown in \textit{Fig.\ref{fig:energy_costs_homo_hertero_64alu_arch}}-\textit{Fig.\ref{fig:normalizedCycleArch64}}, the heterogeneous configuration performs better than the 64-bit homogeneous configuration, with nearly 26\% improvement in performance and nearly 30\% reduction in energy consumption. The 32-bit adder configuration ends up being a close comparison in performance with greatly reduced energy. The 32-bit, and 8-bit configurations share similar properties to the heterogeneous configuration in that they function with improved performance and reduced energy costs over the 64-bit homogeneous configuration. 

In the 32-bit architecture \textit{Fig.\ref{fig:normalizedCycleArch32}}-\textit{Fig.\ref{fig:energy_costs_homo_hertero_32alu_arch}}, for the heterogeneous configuration, we see an improvement in both energy cost and performance over the best-in-performance ALU, 32-bit adder showed a nearly 12\% improvement in performance and nearly 14\% reduction in energy consumption.
Thus the heterogeneous configuration does not trade either energy for performance or vice versa, resulting in a system that grants both improvements.

In regards to potential area costs, though RC adders scale linearly for area costs, it is difficult for us to place an estimate without a full-system integration and analysis of exactly how many adders of each size we will need. What we can conclude though, is that we can add at least 1 of each adder size without exceeding the cost of 2x the area of the largest adder. Based on the analysis of ADD instructions in Dhrystone \textit{Table.\ref{tab1}}, we could estimate that a heterogeneous system may use fewer small adders than large adders due to the lower frequency of small operand commands when compared to large operands at a ratio of approximately 1:2.

Energy constraint 64-bit architecture system \textit{Fig.\ref{fig:energy_costs_homo_hertero_64alu_arch}}, results of energy costs with different sized adders showcase that operation performed on the heterogeneous system at 64-bit incurs fewer energy costs due to the 33\% of instructions, that accounts for less than 16-bit operand size, are assigned to lower bit adder incurs reduced energy costs. 
Similar to a 32-bit system, operating at a reduced energy level of 50\% and further with 25\%, with heterogeneous architecture suggest switching and running all operations on the 8-bit homogeneous system at reduced energy levels that allow complete program execution but with increased latency that is relevant in energy constraint environment.

\begin{figure}[h!]
    \begin{center}    \includegraphics[width=0.5\textwidth]{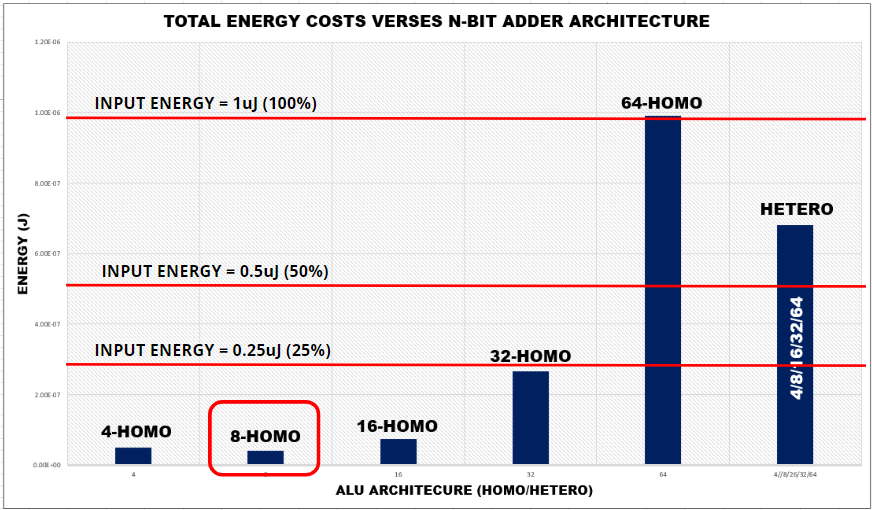}
    \end{center}
    \caption{Energy costs of running the Dhrystone benchmark's ADD operations on N-bit homogeneous architecture and a heterogeneous architecture consisting of adders of each size. This system assumes a max size adder of 64 bits used to handle the largest inputs.}
    \label{fig:energy_costs_homo_hertero_64alu_arch}
\end{figure}

In 64-bit energy constraint system \textit{Fig.\ref{fig:normalizedCycleArch64}}, results of the normalized cycle with different sized adders showcase that 32-bit operation beats out 64-bit performance due to the characteristics of RC-adders resulting in the 64-bit adder having twice the latency of 32-bit. With 64-bit adder accounting for the normalized cycle of 1 compared with a reduction in 26\% cycle with a heterogeneous system.

\begin{figure}[h!]
    \begin{center}    \includegraphics[width=0.5\textwidth]{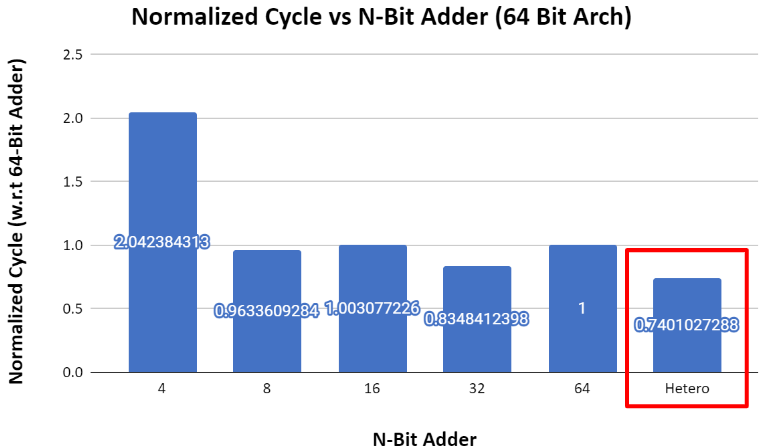}
    \end{center}
    \caption{Average CPI costs, normalized to the 64-bit adder, of running the Dhrystone benchmark's ADD operations on N-bit homogeneous architecture and a heterogeneous architecture consisting of adders of each size. This system assumes a max size adder of 64 bits used to handle the largest inputs.}
    \label{fig:normalizedCycleArch64}
\end{figure}

Energy constraint 32-bit architecture system \textit{Fig.\ref{fig:energy_costs_homo_hertero_32alu_arch}}, results of energy costs with different sized adders showcase that operation performed on heterogeneous system beats out 32-bit energy costs due to the 33\% of instructions, that account for less than 16-bit operand size, are assigned to lower bit adder incurs reduced energy costs. 
In case of available energy costs, reduced to 50\% and further with 25\%, with heterogeneous architecture suggest switching and running all operations on the 8-bit homogeneous system at reduced energy levels that allows complete program execution but with increased latency that is relevant in energy constraint environment. 

\begin{figure}[h!]
    \begin{center}    \includegraphics[width=0.5\textwidth]{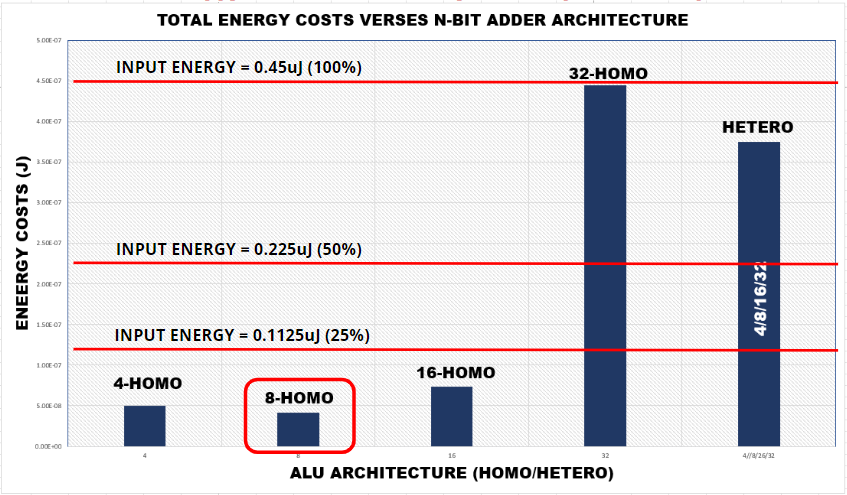}
    \end{center}
    \caption{Energy costs of running the Dhrystone benchmark's ADD operations on N-bit homogeneous architecture and a heterogeneous architecture consisting of adders of each size. This system assumes a max size adder of 32-bit used to handle the largest inputs.}
    \label{fig:energy_costs_homo_hertero_32alu_arch}
\end{figure}

Energy constraint 32-bit architecture system \textit{Fig.\ref{fig:normalizedCycleArch32}}, results of the normalized cycle with different sized adders showcase that heterogeneous system performs with better latency compared to 32-bit operation. The heterogeneous operation, normalized against the best-performing RC-adder results in a reduction of total cycle count of 12\% compared to 32-bit homogeneous architecture running all operations.

\begin{figure}[h!]
    \begin{center}    \includegraphics[width=0.5\textwidth]{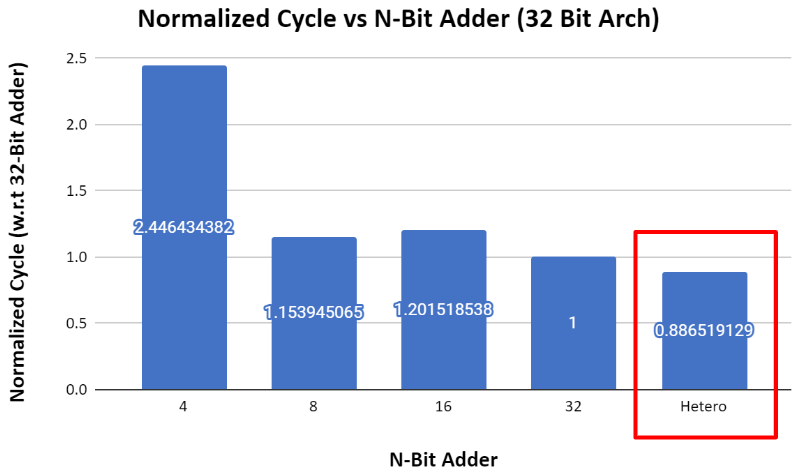}
    \end{center}
    \caption{Average CPI costs, normalized to the 32-bit adder, of running the Dhrystone benchmark's ADD operations on N-bit homogeneous architecture and a heterogeneous architecture consisting of adders of each size. This system assumes a max size adder of 32 bits used to handle the largest inputs.}
    \label{fig:normalizedCycleArch32}
\end{figure}

\section{Future Work}
To further research into heterogeneous ALU systems, there are a couple of topics that should be explored. The evaluation of different ALUs and varied ALU designs would help prove the validity of such a system for various operations. Exploration of different frequencies would show potential gains in various systems. Especially important is system simulator integration. With a system simulation, and accounting for system energy costs as well as pipeline stalls from operations, a more accurate cost and benefit of such an approach could be analyzed. Additionally, a system integration, and defined quantities of ALUs would allow evaluation of additional area cost.

\section{Conclusions}
We believe that heterogeneous ALU systems show promise. It is clear that having available smaller ALUs can allow for the performing of tasks with over an order of magnitude reduced energy costs while maintaining minimally-reduced performance. Furthermore, the heterogeneous ALU system, by budgeting tasks to properly-sized ALUs, is able to not only reduce energy costs of operations but to reduce latency as well. We have yet to integrate our research into a system simulation, which would allow us to properly account for system energy costs and time spent idle, as well as to budget proper quantities of ALUs. However, from our review of the energy and performance characteristics of adders, as well as the frequencies of various-sized ADD operations in the Dhrystone benchmark, we can conclude that there is potential to improve over existing homogeneous ALU systems in both energy costs and performance modestly with 12-26\% performance improvements seen, and simultaneously a 14-30\% reduction in energy costs. We can further conclude that much like heterogeneous CPUs, such a system could be repurposed to serve under energy constraints. The use of such ALU configurations appears especially promising for energy-aware systems, as the potential to reduce energy costs by more than 90\% could allow for extended operation of a system. Much like the governor modes on mainstream heterogeneous CPUs, one could foresee similar modes corresponding to active ALU configurations, chosen to match some particular runtime requirements.

\bibliography{references}
\bibliographystyle{unsrt}

\end{document}